\documentclass{article}
\usepackage[utf8]{inputenc} 
\usepackage[T1]{fontenc}
\usepackage{spconf,amsmath,graphicx}
\usepackage{enumitem}
\setlist{nosep, leftmargin=14pt}
\usepackage{mwe} 
\usepackage{hyperref}       
\usepackage{url}            
\usepackage{booktabs}       
\usepackage{amsfonts}       
\usepackage{nicefrac}       
\usepackage{microtype}      
\usepackage{lipsum}
\usepackage{stackengine}
\usepackage{float}
\usepackage{amssymb}
\usepackage{subfig}
\usepackage{float}
\usepackage{color, colortbl}
\usepackage{rotating}
\usepackage{subfig}
\usepackage{tikz}
\usepackage{xcolor}
\usepackage{animate}
\usepackage{mathrsfs}
\usepackage{multirow}

\usepackage[ruled,vlined]{algorithm2e}

\hypersetup{
    colorlinks,
    linkcolor={red!50!black},
    citecolor={blue!80!black},
    urlcolor={blue!80!black}
}
\usepackage{setspace}
\title{Deep Deterministic Nonlinear ICA \\ via Total Correlation Minimization with Matrix-Based Entropy Functional}
\name{Qiang Li$^{1}$, Shujian Yu$^{2}$, Liang Ma$^{1}$, Chen Ma$^{3}$, \\ Jingyu Liu$^{1,4}$, Tülay Adali$^{5}$, Fellow, IEEE, Vince D. Calhoun$^{1,4}$, Fellow, IEEE}
\address{$^{1}$Tri-Institutional Center for Translational Research in Neuroimaging and Data Science (TReNDS), \\
Georgia State, Georgia Tech, and Emory University, Atlanta, GA, United States \\
$^{2}$Department of Computer Science, Vrije Universiteit Amsterdam, The Netherlands \\
$^{3}$School of Research Center of Satellite Technology, Harbin Institute of Technology, Harbin, China \\
$^{4}$Department of Computer Science, Georgia State University, Atlanta, GA, United States \\
$^{5}$Department of Computer Science and Electrical Engineering,  \\
University of Maryland, Baltimore County, Baltimore, MD, United States}

\begin{document}
\maketitle
\begin{abstract}
Blind source separation, particularly through independent component analysis (ICA), is widely utilized across various signal processing domains for disentangling underlying components from observed mixed signals, owing to its fully data-driven nature that minimizes reliance on prior assumptions. However, conventional ICA methods rely on an assumption of linear mixing, limiting their ability to capture complex nonlinear relationships and to maintain robustness in noisy environments. In this work, we present deep deterministic nonlinear independent component analysis (DDICA), a novel deep neural network-based framework designed to address these limitations. DDICA leverages a matrix-based Rényi’s $\alpha$-order entropy function to directly optimize the independence criterion via stochastic gradient descent, bypassing the need for variational approximations or adversarial schemes. This results in a streamlined training process and improved resilience to noise. We validated the effectiveness and generalizability of DDICA across a range of applications, including simulated signal mixtures, hyperspectral image unmixing, modeling of primary visual receptive fields, and resting-state functional magnetic resonance imaging (fMRI) data analysis. Experimental results demonstrate that DDICA effectively separates independent components with high accuracy across a range of applications. These findings suggest that DDICA offers a robust and versatile solution for blind source separation in diverse signal processing tasks.
\end{abstract}


\section{Introduction}
\label{sec:intro}
\hspace{1.2em} Independent component analysis (ICA), as a blind source separation (BSS) method, has been widely used in various types of signal analyses to extract statistically independent source signals from observed mixtures~\cite{comon10,hyvarinen1999fast,bell1995information}. 

In ICA, the observed data are modeled as linear mixtures of unknown source signals, and ICA can exploit different statistical properties to achieve separation, including higher-order statistics, second-order statistics, temporal structure, or other forms of diversity~\cite{hyvarinen1999fast,bell1995information,Adali14IEE}. The most well-known ICA methods based on higher-order statistics fundamentally rely on the non-Gaussianity of the source signals, with prominent examples including FastICA~\cite{hyvarinen1999fast} and Infomax~\cite{bell1995information}. These algorithms seek to approximate statistical independence by either maximizing the non-Gaussianity of the extracted components (often using approximations of negentropy or kurtosis) or, equivalently, by minimizing the mutual information between them. Higher-order statistics-based ICA has proven highly robust in uncovering hidden structures and is widely applied across diverse fields, including biomedical signal processing for data like fMRI and EEG~\cite{Calhoun2009ARO, Yuhui20, Correa05gift}, audio source separation~\cite{Makino04, Sattar01}, and natural image analysis for extracting essential, statistically independent features~\cite{Hateren99,Hyvarinen2009}.

Over the decades, linear ICA has found wide-ranging applications across diverse domains. However, it has also faced criticism for its limited effectiveness in handling real-world data characterized by nonlinear mixing. To address this limitation, numerous approaches have been proposed to extend ICA into the nonlinear domain~\cite{hyvarinen1999nonlinear, Aapo23PATTERN}. The most direct generalization involves replacing the linear mixing matrix $\mathbf{A}$ with a nonlinear, typically smooth and invertible function $\mathbf{f}(\cdot)$, thereby transforming the observed data into $\mathbf{x} = \mathbf{f}(\mathbf{s})$. The objective then becomes to recover the inverse transformation $\mathbf{f}^{-1}(\cdot)$ in order to retrieve the latent sources~\cite{Sorrenson2020Disentanglement,khemakhem20a,Almeida02}.

However, it is well-established that the unsupervised learning of identifiable nonlinear ICA is theoretically impossible without introducing additional structural constraints or auxiliary information. In the absence of such constraints, the estimated sources $\hat{\mathbf{s}}$ may appear statistically independent but are inherently non-unique, rendering the problem fundamentally ill-posed.

In this work, we introduced deep deterministic nonlinear independent component analysis (DDICA), which uses a novel estimation approach to solve the problem. DDICA employs a single feed-forward autoencoder as the unmixing-mixing function and incorporates a backpropagation-friendly whitening layer before the bottleneck layer to handle nonlinear scenarios~\cite{Li2022DeepDI}. The neural network is trained non-parametrically using the total correlation~\cite{watanabe1960information}, with the recently proposed matrix-based Rényi's $\alpha$-order entropy~\cite{Giraldo2014,Yu2019,Dong23}. DDICA can be trained directly using stochastic gradient descent and its variants, eliminating the need for variational approximations. To validate DDICA, we applied it to a variety of tasks. First, we conducted a simulation experiment, where DDICA was compared with existing linear and nonlinear ICA algorithms, demonstrating its competitive performance. We then applied DDICA to hyperspectral image unmixing, utilized it for modeling primary visual receptive fields, and tested it on resting-state fMRI data. These diverse applications highlight DDICA’s generalizability and effectiveness across different domains in signal processing.

\section{Background}
\subsection{Linear and Nonlinear Independent Component Analysis}
\subsubsection{Linear ICA} 
\hspace{1.2em} Linear ICA assumes that the observed set of  \(d\)-dimensional vectors \(\mathbf{x} = [x_1, x_2, \cdots, x_d] ^T\in \mathbb{R}^{d \times m}\) is generated by a set of \(n\) latent independent components, \(\mathbf{s} = [s_1, s_2, \cdots, s_n]^T\in \mathbb{R}^{n \times m}\), according to the model:
\[
\mathbf{x} = \mathbf{A}\mathbf{s},
\]
where \(\mathbf{A}\) is a mixing matrix, \( \in\mathbb{R}^{d \times m}\). The goal of linear ICA is to recover the inverse of the mixing matrix \(\mathbf{A}^{-1}\), characterized by an unmixing matrix \(\mathbf{W}\), and to identify the independent components \(\mathbf{s} = [s_1, s_2, \cdots, s_n]\) based solely on the observed data \(\mathbf{x}\):
\[
\mathbf{\tilde{s}} = \mathbf{W}\mathbf{x}.
\]

Linear ICA can be understood through two main dominant strategies. The first strategy aims to maximize the non-Gaussianity of the estimated sources, drawing on the central limit theorem~\cite{lecam1986clt}. It typically employs measures such as kurtosis and negentropy, based on the idea that observed mixed signals are generally more Gaussian than the underlying independent components~\cite{pati2021independent}. The second strategy focuses on minimizing the mutual information (\texttt{MI}) among the estimated sources, where \texttt{MI} is defined as the Kullback-Leibler (\texttt{KL}) divergence between the joint distribution of the sources and the product of their marginal distributions~\cite{bell1995information}.

\subsubsection{Nonlinear ICA} 
\hspace{1.2em} The most straightforward extension of ICA to the nonlinear setting involves replacing the linear mixing matrix \(\mathbf{A}\) with an invertible mixing function \(\mathbf{f}(\cdot)\), leading to the model:
\[
\mathbf{x} = \mathbf{f}(\mathbf{s}),
\]
where \(\mathbf{x}\) is the observed data and \(\mathbf{s}\) represents the latent sources. In this case, the goal is to recover the inverse function \(\mathbf{f}^{-1}\), which is characterized by an unmixing function \(\mathbf{g}\), such that:
\[
\mathbf{\tilde{s}} = \mathbf{g}(\mathbf{x}).
\]
\hspace{1.2em} Identifiability in the nonlinear ICA framework cannot always be guaranteed. This challenge arises because, without constraints on the space of mixing functions, there can be an infinite number of possible solutions~\cite{hyvarinen1999nonlinear}. However, recent advances using deep neural networks (DNNs) have led to identifiable results by leveraging the temporal patterns present in the raw observations~\cite{belghazi2018mutual}. These developments help address the issues of identifiability in nonlinear ICA by incorporating learned representations from data.

\subsection{Mutual Information, Total Correlation}
\subsubsection{Mutual Information}
\hspace{1.2em} The \texttt{MI} measures the amount of information obtained about one variable through another variable~\cite{cover1991information}. It quantifies the degree of dependence between two variables. The \( I(\mathbf{x}_1; \mathbf{x}_2) \) between two variables \( \mathbf{x}_1 \) and \( \mathbf{x}_2 \) can be defined in terms of entropy as:
\begin{equation}
\texttt{I}(\mathbf{x}_1; \mathbf{x}_2) = H(\mathbf{x}_1) + H(\mathbf{x}_2) - H(\mathbf{x}_1, \mathbf{x}_2).    
\end{equation}

\noindent where \( H(\mathbf{x}_1) \) is the entropy of the variable \(\mathbf{x}_1 \), representing the average uncertainty or the amount of information contained in \( \mathbf{x}_1 \), and \( H(\mathbf{x}_1, \mathbf{x}_2) \) is the joint entropy of \( x_1 \) and \( x_2 \), representing the average uncertainty of the combined system consisting of both variables \( \mathbf{x}_1 \) and \( \mathbf{x}_2 \). If two variables are independent, the \texttt{MI} will be zero.

\subsubsection{Total Correlation}
\hspace{1.2em} Total Correlation (\texttt{TC}), also known as multivariate mutual information, characterizes the dependence among $n$ variables. It can be seen as a non-negative generalization of \texttt{MI}, extending the concept from pairs of variables to multiple variables.

\noindent Let \texttt{TC}, as defined by~\cite{watanabe1960information}, be denoted as:
\begin{equation}\label{eq.tc}
\small
\begin{split}
   &\texttt{TC}\left(\mathbf{x}_1, \ldots, \mathbf{x}_{n}\right) \equiv \sum_{i=1}^n H\left(\mathbf{x}_{i}\right)-H\left(\mathbf{x}_1, \ldots, \mathbf{x}_{n}\right) \\ 
   &=D_{KL}\left(P\left(\mathbf{x}_1, \ldots, \mathbf{x}{n}\right) \| \prod_{i=1}^{n} p\left(\mathbf{x}_{i}\right)\right).
\end{split}
\end{equation}

\noindent where $H(\mathbf{x}_1 )$ denotes marginal entropy, and $H(\mathbf{x}, \cdots, \mathbf{x}_{n})$ represents joint entropy. From definitions in (\ref{eq.tc}), if all variables are independent, the \texttt{TC} will be zero. When \( n = 2 \), \texttt{TC} equals to \texttt{MI}. Therefore, multivariate dependencies can be effectively quantified using \texttt{TC}, which addresses the limitations of \texttt{MI} that captures only pairwise relationships~\cite{li2022functionalen,QiangVis}. It has also been recently applied for disentangled representation learning~\cite{chen2018isolating}, understanding the learning dynamics of DNNs~\cite{yu2021measuring}, and estimating high-order functional connectivity in human brain~\cite{gatica2021high,li2022functional,li2023aberrant}. 

\section{METHODOLOGY}
\subsection{Deep Deterministic Nonlinear Independent Component Analysis}
\hspace{1.2em} In deep deterministic nonlinear independent component analysis (DDICA), we use \( \text{DNN} \, g_\theta \), parameterized by \( \theta \), as a nonlinear unmixing matrix, as illustrated in Fig.\ref{fig:1}. Our objective is to minimize the total dependence across all estimated sources $\left[\tilde{s}_1,\tilde{s}_2,\cdots,\tilde{s}_p\right]$. Here we define the total dependence for \(\left[\tilde{s}_1, \tilde{s}_2, \ldots, \tilde{s}_p\right]\) as the Kullback–Leibler (\texttt{KL}) divergence from the joint distribution $P(\tilde{s}_1,\tilde{s}_2,\cdots,\tilde{s}_p)$ to the product of marginal distributions $P(\tilde{s}_1)P(\tilde{s}_2)\cdots P(\tilde{s}_p)$: 

\begin{equation}
\begin{aligned}
    \texttt{TC} & = D_{KL}\left[P(\tilde{s}_1,\tilde{s}_2,\cdots,\tilde{s}_p);P(\tilde{s}_1)P(\tilde{s}_2)\cdots P(\tilde{s}_p)\right],\\
    & = \left[\sum_{i=1}^p H(\tilde{s}_i)\right] - H(\tilde{s}_1,\tilde{s}_2,\cdots,\tilde{s}_p),
\end{aligned}
\label{eq:total_correlation}
\end{equation}

\noindent where $H(\tilde{s}_i)$ is the entropy of the $i$-th source, $H(\tilde{s}_1,\tilde{s}_2,\cdots,\tilde{s}_p)$ is the joint entropy for $\left[\tilde{s}_1,\tilde{s}_2,\cdots,\tilde{s}_p\right]$. To make it more applicable to DNNs, an alternative way to define the total dependence is given by the following formula:

\begin{equation}
  D_{KL}=\sum_{i=1}^p I\left(\tilde{s}_i ; \tilde{\mathbf{s}}_{-i}\right),  
\end{equation}

\noindent where \( I \) denotes mutual information and \( \tilde{\mathbf{s}}_{-i} \) represents the set of all sources except the \( i \)-th source. In DNNs, to make the training process more practical and straightforward, we compute the entropy and joint entropy terms in (\ref{eq:total_correlation}) directly from the data using the matrix-based R\'enyi \(\alpha\)-order entropy functional, without relying on any variational approximations or distributional assumptions.

Suppose there are \( N \) samples\footnote{Here, \( N \) can be considered analogous to the batch size in neural network training.} for the \( i \)-th predicted source, denoted as \( \tilde{s}_{i} = [\tilde{s}_{i}^1, \tilde{s}_{i}^2, \ldots, \tilde{s}_{i}^N] \), where the subscript denotes the view index and the superscript denotes the sample index. A Gram matrix \( K_i \in \mathbb{R}^{N \times N} \) can be computed with entries given by \( K_i(n,m) = \kappa(\tilde{s}_{i}^n, \tilde{s}_{i}^m) \), where \( \kappa \) is an infinitely divisible kernel~\cite{bhatia2006infinitely}, which is usually assumed to be Gaussian. The entropy of \( \tilde{s}_{i} \) can be expressed as~\cite{giraldo2014measures}:

\begin{figure}[!ht]
    \centering
    \includegraphics[width=0.5\textwidth, height=4.5cm]{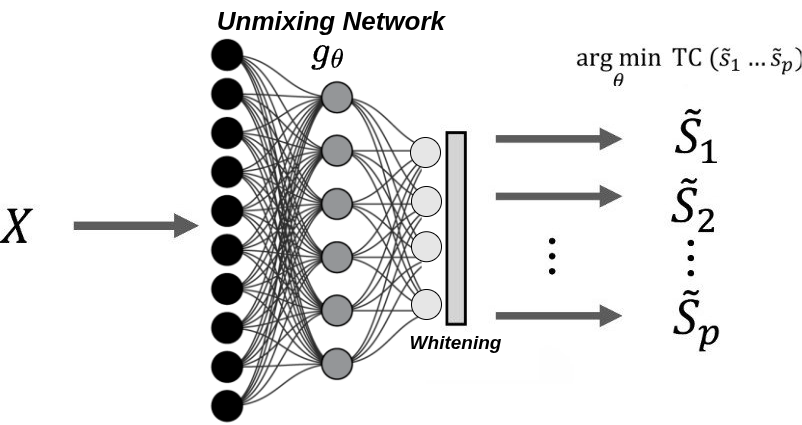}
    \caption{\textbf{The architecture of DDICA.} For DDICA, we aim to minimize the overall dependency among all predicted sources \(\{\tilde{s}_i\}_{i=1}^p\) by optimizing the objective function \(\underset{\theta}{\arg\min}~ \texttt{TC}\left(\tilde{s}_{1},\cdots,\tilde{s}_{p}\right)\). Then, a differentiable whitening layer is applied before the output to separate the sources.}
    \label{fig:1}
\end{figure}

\begin{equation}\label{eq:renyi_entropy}
H_{\alpha}(\tilde{s}_{i}) = H_{\alpha }(A_i) = \frac{1}{1-\alpha }\log_{2}[\sum_{n=1}^{N}\lambda _{n}(A_i)^{\alpha}],
\end{equation}
where $A_i=K_i/tr(K_i)$ is the normalized Gram matrix and $\lambda _{n}(A_i)$ denotes $n$-th eigenvalue of $A_i$. In our study, the order \(\alpha\) is set to 0.75, and the kernel width \(\sigma\) is 0.1584, as determined using Silverman’s rule~\cite{silverman1986density}.

Further, the joint entropy for \(\{\tilde{s}_i\}_{i=1}^p\) is defined as~\cite{Yu2019}:

\begin{equation}\label{eq:renyi_joint_entropy}
\begin{aligned}    
&H_{\alpha}(\tilde{s}_1, \tilde{s}_2, \ldots, \tilde{s}_p) = H_{\alpha}\left(\frac{A_1 \circ A_2 \circ \cdots \circ A_p}{\text{tr}(A_1 \circ A_2 \circ \cdots \circ A_p)}\right) \\
& \frac{1}{1-\alpha} \log \left(\operatorname{tr}\left(\left(\frac{A^1 \circ A^2 \circ \cdots \circ A^n}{\operatorname{tr}\left(A^1 \circ A^2 \circ \cdots \circ A^n\right)}\right)^\alpha\right)\right),
\end{aligned}
\end{equation}

The \texttt{TC} in (\ref{eq:total_correlation}) can be directly obtained using (\ref{eq:renyi_entropy}) and (\ref{eq:renyi_joint_entropy}). This leads to,

\begin{equation}
\begin{aligned}
&\texttt{TC}= \left[\sum_{i=1}^p H(\tilde{s}_i)\right] - H(\tilde{s}_1,\tilde{s}_2,\cdots,\tilde{s}_p) \\
& =\left[\sum_{i=1}^p \frac{1}{1-\alpha} \log \left(\operatorname{tr}\left(\frac{A^i}{\operatorname{tr}\left(A^i\right)}\right)^\alpha\right)\right] -\\
&\frac{1}{1-\alpha} \log \left(\operatorname{tr}\left(\left(\frac{A^1 \circ A^2 \circ \cdots \circ A^n}{\operatorname{tr}\left(A^1 \circ A^2 \circ \cdots \circ A^n\right)}\right)^\alpha\right)\right).
\end{aligned}
\end{equation}

The differentiability of this matrix-based R\'enyi entropy estimator has been both theoretically demonstrated and empirically confirmed in~\cite{yu2021measuring}. In practical applications, automatic singular value decomposition is integrated into major deep learning frameworks such as PyTorch. Consequently, our DDICA enjoys a straightforward and manageable objective function, \(\min \texttt{TC}\), which can be efficiently optimized using singular value decomposition or its variants. 

In scenarios involving nonlinearity, it is important to incorporate a differentiable whitening layer followed by unmixing networks within the model. The goal of the whitening operator is to ensure that the latent space is decorrelated and standardized. In nonlinear settings, this requires a differentiable whitening approximator to be applied directly to the predicted source signals. To achieve this, we implement power iteration~\cite{vonmises1929gleichungsaufloesung}, a widely used method for computing a differentiable eigendecomposition. Given the covariance matrix of the predicted sources, 

\begin{equation}
\mathbf{C} = \mathbb{E}[\mathbf{g}(\mathbf{x}) \mathbf{g}(\mathbf{x})^\top] = \mathbf{U} \mathbf{D} \mathbf{U}^\top,
\end{equation}
we iteratively approximate the dominant eigenvector \( \mathbf{u} \) and its corresponding eigenvalue \( \lambda \) using the update:
\begin{equation}
    \mathbf{u}^{(i+1)} = \frac{\mathbf{C} \mathbf{u}^{(i)}}{\lambda^{(i)}}, \quad \text{with} \quad \lambda^{(i)} = \left\| \mathbf{C} \mathbf{u}^{(i)} \right\|,
\end{equation}
initialized with a random vector \( \mathbf{u}^{(0)} \in \mathbb{R}^{d} \). After estimating each eigenvector–eigenvalue pair, we remove its spectral contribution from the covariance matrix:
\begin{equation}
    \mathbf{C} \gets \mathbf{C} - \lambda \mathbf{u} \mathbf{u}^\top.
\end{equation}
Repeating this process allows extraction of all eigenvector–eigenvalue pairs \( \{ \mathbf{u}_j, \lambda_j \} \) in descending order of eigenvalues.

Finally, the whitening matrix is constructed from the full set of eigenpairs:
\begin{equation}
    \mathbf{U} \mathbf{D}^{-1/2} \mathbf{U}^\top = \sum_{j=1}^{d} \frac{1}{\sqrt{\lambda_j}} \mathbf{u}_j \mathbf{u}_j^\top.
\end{equation}

Various stopping criteria can be employed to assess convergence; however, for simplicity, we adopt a fixed number of power iterations in this study. This decomposition is fully based on differentiable operations and is integrated as a key component of the nonlinear DDICA framework, as illustrated in Fig.\ref{fig:1}.

In DDICA, we utilized a deep neural network comprising 9 fully connected layers, followed by a differentiable whitening layer, to analyze and extract latent components. For training the model, we used a learning rate of \( 0.0001 \) to control the step size during optimization. In addition, the Adam optimizer was employed to efficiently adjust the model parameters, leveraging its adaptive learning rate capabilities to improve convergence and performance.

\begin{figure*}[!htbp]
    \centering
    \includegraphics[width=\textwidth,height=20cm]{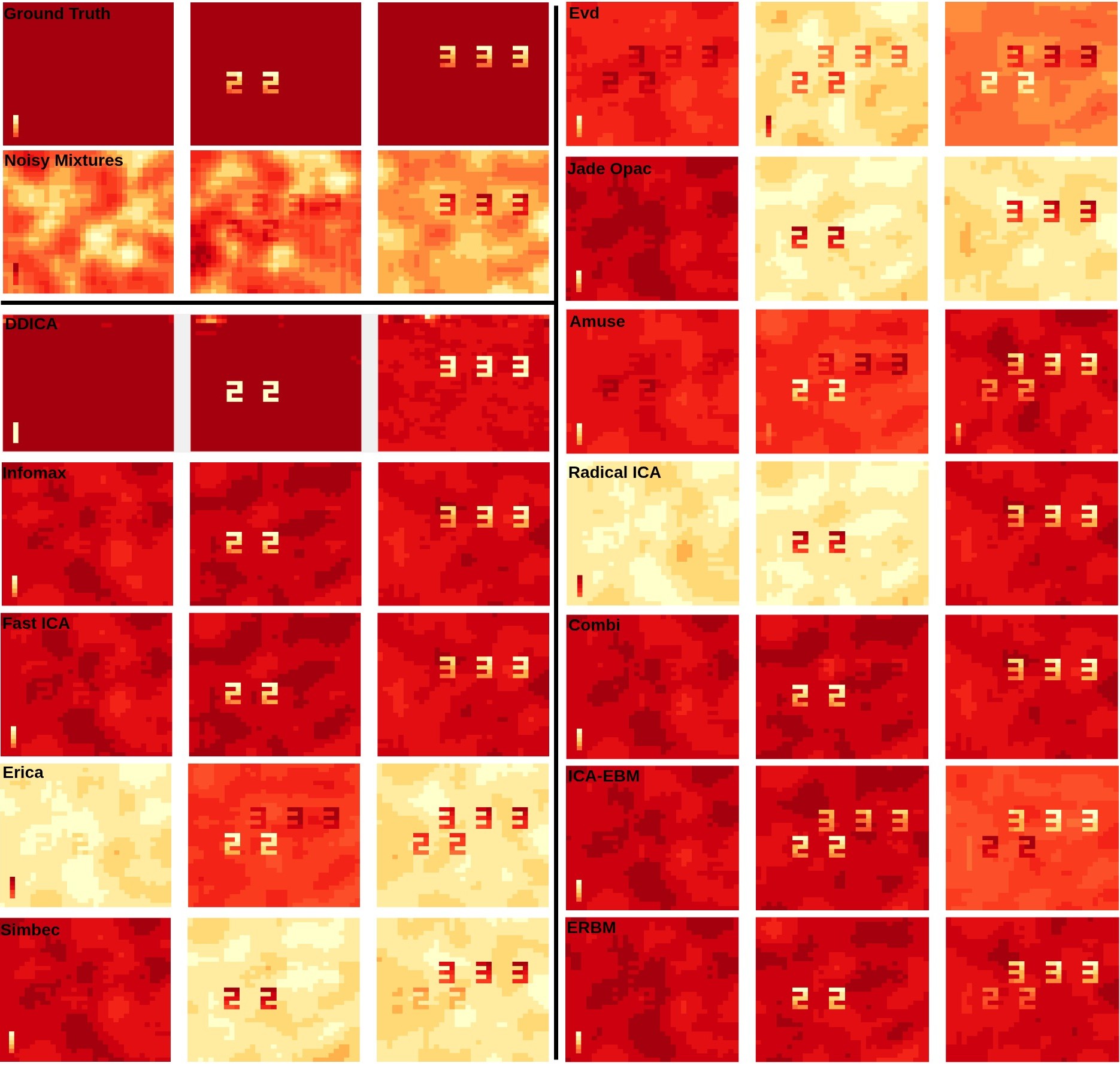}
    \caption{\textbf{Comparison of DDICA with 11 other linear ICA algorithms.} Simulated ground-truth sources (sources 1, 2, and 3) and their corresponding noisy mixtures are shown for reference. To evaluate the performance of DDICA, we compared its results with those from 11 established linear ICA algorithms: Infomax, FastICA, Erica, Simbec, Evd, Jade Opac, Amuse, Radical ICA, Combi, ICA-EBM, and ERBM. The results highlight the ability of DDICA to recover sources with high accuracy and robustness compared to other linear ICA approaches.}
    \label{fig:simu}
\end{figure*}

\begin{figure*}[!ht]
    \centering
    \includegraphics[width=\textwidth,height=4.3cm]{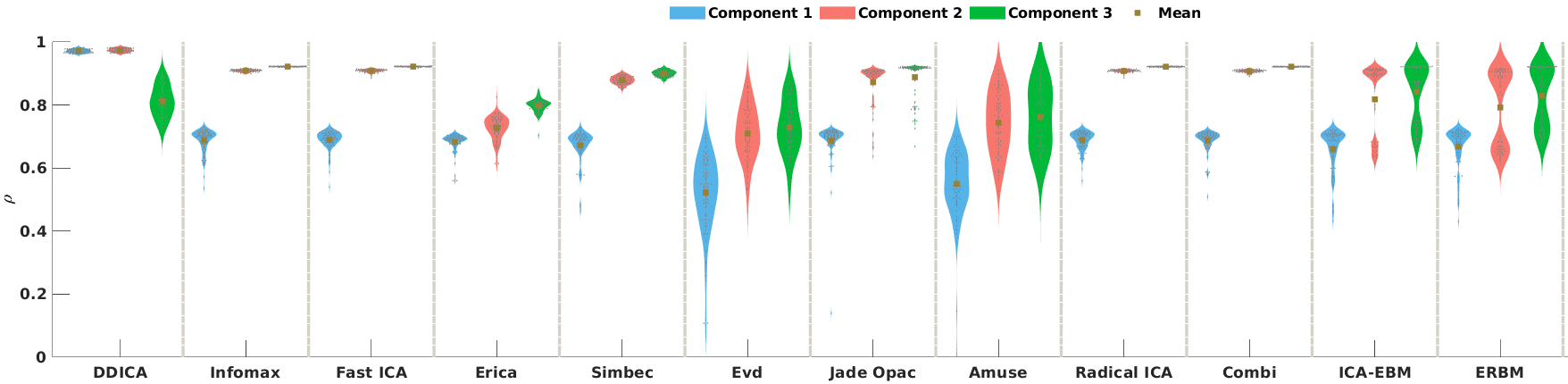}
    \caption{\textbf{Quantitative evaluation of DDICA compared to 11 other linear ICA algorithms.} To assess the performance of DDICA, we conducted 90 repeated simulation trials. In each trial, spatial correlation was measured between the estimated and ground truth components across three source signals. The average results for DDICA and 11 other linear ICA algorithms (Infomax, FastICA, Erica, Simbec, Evd, Jade Opac, Amuse, Radical ICA, Combi, ICA-EBM, and ERBM) are presented. This evaluation reveals that DDICA provides the highest performance for sources 1 and 2 and the highest overall performance across all methods and highlights the consistency and accuracy of DDICA in recovering independent sources across repeated experiments.}
    \label{fig:simun}
\end{figure*}

\begin{table*}[!htbp]
\centering
\caption{\textbf{Benchmarking DDICA against 11 widely used linear ICA algorithms.} Here presents the average spatial similarity (along with standard deviation) between the estimated and ground truth components across 90 repeated trials for DDICA and 11 other linear ICA algorithms. An ICA algorithm is considered deterministic if it consistently produces the exact same output when given identical input data and parameters. In contrast, a nondeterministic algorithm may yield varying results across runs due to internal sources of randomness.}
\scalebox{0.89}{
\begin{tabular}{@{}llllc@{}}
\toprule
Class                                                                                                                                                                                                             & Algorithms                       & References                                                                                                    & Types                                  & Spatial Similarities  \\ \midrule
\multicolumn{1}{|l|}{\multirow{4}{*}{\begin{tabular}[c]{@{}l@{}}Maximum  Likelihood Based \\ (Estimate at most one Gaussian)\end{tabular}}}                                                                        & \multicolumn{1}{l|}{Infomax}     & \multicolumn{1}{l|}{\begin{tabular}[c]{@{}l@{}}Bell et al., 1995~\cite{bell1995information}\\ Correa et al., 2007~\cite{Correa2007PerformanceOB}\end{tabular}}         & \multicolumn{1}{l|}{non deterministic} & \multicolumn{1}{c|}{$0.84\pm0.01$} \\ \cmidrule(l){2-5} 
\multicolumn{1}{|l|}{}                                                                                                                                                                                            & \multicolumn{1}{l|}{FastICA}     & \multicolumn{1}{l|}{Hyvärinen et al., 1999~\cite{hyvarinen1999fast}}                                                                   & \multicolumn{1}{l|}{non deterministic} & \multicolumn{1}{c|}{$0.84\pm0.02$} \\ \cmidrule(l){2-5} 
\multicolumn{1}{|l|}{}                                                                                                                                                                                            & \multicolumn{1}{l|}{Radical ICA} & \multicolumn{1}{l|}{\begin{tabular}[c]{@{}l@{}}Learned-Miller et al., 2003~\cite{LearnedMiller03} \end{tabular}} & \multicolumn{1}{l|}{deterministic}     & \multicolumn{1}{c|}{$0.74\pm0.03$} \\ \cmidrule(l){2-5} 
\multicolumn{1}{|l|}{}                                                                                                                                                                                            & \multicolumn{1}{l|}{ICA-EBM}     & \multicolumn{1}{l|}{\begin{tabular}[c]{@{}l@{}}Li et al., 2010~\cite{XiLin10} \end{tabular}}               & \multicolumn{1}{l|}{non deterministic} & \multicolumn{1}{c|}{$0.82\pm0.02$} \\ \midrule
\multicolumn{1}{|l|}{\multirow{4}{*}{\begin{tabular}[c]{@{}l@{}}Maximum Likelihood Based\\ (Separates Gaussians with\\ different sample dependence \\ structure, i.e., autocorrelation\\ matrices)\end{tabular}}} & \multicolumn{1}{l|}{Evd}         & \multicolumn{1}{l|}{Georgiev et al., 2001~\cite{georgiev2001blind}}                                                                    & \multicolumn{1}{l|}{non deterministic} & \multicolumn{1}{c|}{$0.65\pm 0.10$} \\ \cmidrule(l){2-5} 
\multicolumn{1}{|l|}{}                                                                                                                                                                                            & \multicolumn{1}{l|}{ERBM}        & \multicolumn{1}{l|}{\begin{tabular}[c]{@{}l@{}}Fu et al., 2015~\cite{Fu2015sp}\end{tabular}}       & \multicolumn{1}{l|}{non deterministic} & \multicolumn{1}{c|}{$0.82\pm0.06$} \\ \cmidrule(l){2-5} 
\multicolumn{1}{|l|}{}                                                                                                                                                                                            & \multicolumn{1}{l|}{Amuse}       & \multicolumn{1}{l|}{\begin{tabular}[c]{@{}l@{}}Tong et al., 1990, 1991~\cite{tong1990amuse,tong1991amuse}\end{tabular}}  & \multicolumn{1}{l|}{deterministic}     & \multicolumn{1}{c|}{$0.69\pm0.10$} \\ \cmidrule(l){2-5} 
\multicolumn{1}{|l|}{}                                                                                                                                                                                            & \multicolumn{1}{l|}{Combi}       & \multicolumn{1}{l|}{\begin{tabular}[c]{@{}l@{}}Tichavsky et al., 2006, 2011~\cite{Tichavsky06,Tichavsky11}\end{tabular}}      & \multicolumn{1}{l|}{non deterministic} & \multicolumn{1}{c|}{$0.84\pm0.01$} \\ \midrule
\multicolumn{1}{|l|}{\multirow{3}{*}{\begin{tabular}[c]{@{}l@{}}Cumulant-based (Estimate at most\\ one Gaussian)\end{tabular}}}                                                                                   & \multicolumn{1}{l|}{Simbec}      & \multicolumn{1}{l|}{Cruces et al., 2001~\cite{Cruces01}}                                                                      & \multicolumn{1}{l|}{deterministic}     & \multicolumn{1}{c|}{$0.84\pm0.01$} \\ \cmidrule(l){2-5} 
\multicolumn{1}{|l|}{}                                                                                                                                                                                            & \multicolumn{1}{l|}{Erica}       & \multicolumn{1}{l|}{Cruces et al., 2002~\cite{CRUCES200287}}                                                                      & \multicolumn{1}{l|}{deterministic}     & \multicolumn{1}{c|}{$0.77\pm0.09$} \\ \cmidrule(l){2-5} 
\multicolumn{1}{|l|}{}                                                                                                                                                                                            & \multicolumn{1}{l|}{Jade Opac}   & \multicolumn{1}{l|}{\begin{tabular}[c]{@{}l@{}}Cardoso et al., 1993~\cite{cardoso1993blind}\end{tabular}}        & \multicolumn{1}{l|}{deterministic}     & \multicolumn{1}{c|}{$0.76\pm0.09$} \\ \midrule
Total Correlation-based (Deep neural network)                                                                                                                                                                                           & DDICA                            & Current                                                                                                       & deterministic                          & \multicolumn{1}{c}{$0.89\pm0.01$}  \\ \bottomrule
\end{tabular}}
\label{tab}
\end{table*}

\begin{figure}[!ht]
    \centering
    \includegraphics[width=0.5\textwidth,height=14.7cm]{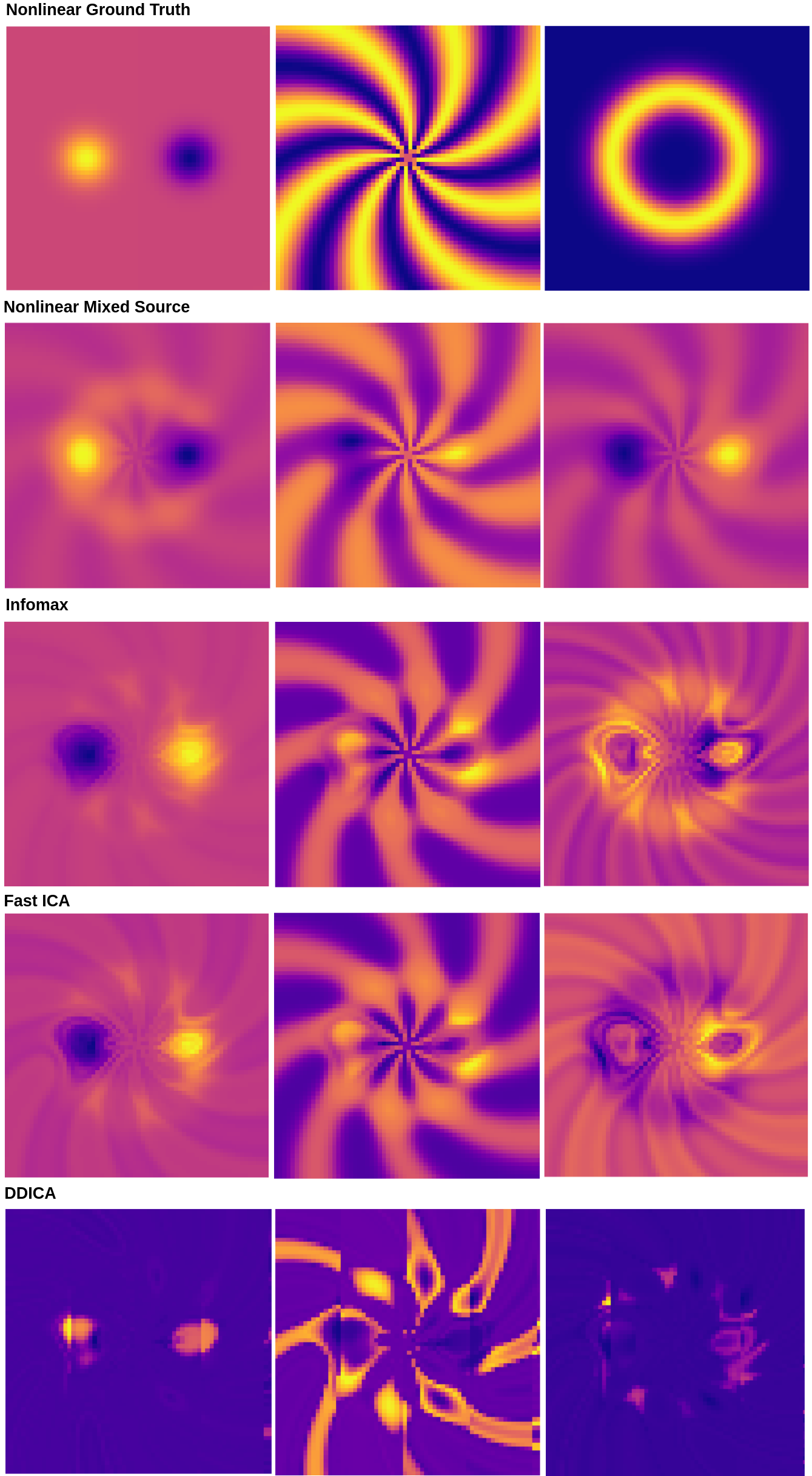}
    \caption{\textbf{Decomposing stronger nonlinear mixed sources.} Strongly nonlinear signals such as blobs, spiral waves, and circular patterns were mixed to create extremely nonlinear mixed sources. The performance of top ICA algorithms, Infomax and FastICA, was then tested and compared against DDICA.}
    \label{fig:n2}
\end{figure}


\section{EXPERIMENTAL RESULTS}
\label{sec:res}
\subsection{Datasets and Preprocessing}
\hspace{1.2em} \textbf{Synthesis Data I:} We generated three ground-truth source signals, each defined as a \(33 \times 33\) image and repeated over 50 time frames (\(t = 50\)). These sources correspond to three independent components, with pixel activations shaped to resemble the patterns “1”, “2 2”, and “3 3 3”. Pixel intensities within these shapes range from 0.5 to 1, while all other pixels are set to zero.

To simulate realistic observations, we added structured noise to the source signals in two stages, as described in~\cite{risk2021simultaneous,wang2024sparse}. Initially, at time point \(t = 1\), a Gaussian random field was generated with a standard deviation of 1 and spatial smoothness controlled by a Gaussian kernel with full width at half maximum (FWHM) of 6. For subsequent time points (\(t = 2\) to \(50\)), noise was generated using a first-order autoregressive (AR(1)) model: the previous noise frame was multiplied by 0.47 and added to a newly sampled independent Gaussian random field, again with FWHM = 6.

To control the signal-to-noise ratio (SNR), the variance \(\sigma^2\) of the Gaussian noise \(\mathbf{N}\) was adjusted. Let \(\lambda_1, \dots, \lambda_Q\) denote the nonzero eigenvalues of the covariance matrix of the source mixture \(\mathbf{S}_M\). The SNR was defined as:

\[
\text{SNR} = \frac{1}{T \cdot \sigma^2} \sum_{i=1}^{Q} \lambda_i,
\]

\noindent and was fixed at a value of 0.4 for all simulations.

\textbf{Synthesis Data II:} We generate synthetic 2D nonlinear mixtures of spatial signals, specifically designed to evaluate the performance of DDICA algorithms in scenarios involving strongly nonlinear mixing. It constructs a three-source dataset \( S(x, y, i) \), where each source represents a distinct spatial pattern: a circular ring defined by a radial Gaussian centered at \( r = 0.5 \), a sinusoidal spiral dependent on both angular (\( \theta \)) and radial (\( r \)) components, and a bipolar Gaussian configuration formed by two symmetrically placed blobs along the horizontal axis. Each source is normalized to have zero mean and unit variance to ensure statistical consistency across components. A random linear mixing matrix \( \mathbf{A} \in \mathbb{R}^{3 \times 3} \) is applied to the vectorized sources to simulate instantaneous linear mixing. Subsequently, a controlled nonlinear transformation is introduced using both hyperbolic tangent and sinusoidal functions scaled by a nonlinearity coefficient \( \alpha \), such that the final mixed signals are given by \(\mathbf{X}_{\text{n}} = \mathbf{A} \mathbf{S} + \alpha \tanh(\mathbf{A} \mathbf{S}) + \alpha^2 \sin(\mathbf{A} \mathbf{S}) \), where \( \alpha \in [0,1] \) controls the nonlinearity level. This structure creates a challenging benchmark for evaluating nonlinear ICA performance under realistic signal deformation.

\textbf{Synthesis Data III:} We generate samples in a two-dimensional space, each associated with one of five distinct cluster labels (Ground Truth). The cluster centers are independently sampled from a uniform distribution over the interval $[-5, 5]$, while their variances are drawn from a uniform distribution between 0.5 and 3. To construct a ten-dimensional latent space, we append eight additional dimensions of independent Gaussian noise to this 2D data. These noise components are scaled by a factor of 0.01, ensuring they remain negligible relative to the informative dimensions, which are limited to the first two.

\textbf{Hyperspectral Images:}
We utilized a real-world hyperspectral dataset, namely Urban~\cite{kalman1997classification}. The dataset consists of a $307 \times 307$ pixel image, with each pixel corresponding to a $2 \times 2\text{m}^2$ area. It contains 210 spectral bands, ranging from 400nm to 2500nm, offering a spectral resolution of 10nm. After removing channels 1-4, 76, 87, 101-111, 136-153, and 198-210 due to dense water vapor and atmospheric effects, 162 spectral bands remain. 

\textbf{Natural Images:} In this study, both grayscale and chromatic images are used as natural image datasets~\cite{urs2022unsupervised}. Image patches are extracted from the input images and flattened into 1D vectors to create a matrix of samples × features. For our experiments, we use 100K and 500K samples for each modality. The patch sizes are 8x8 pixels and 16x16 pixels for both grayscale and color images, with color images having channel information (e.g., 8x8x3 and 16x16x3 for color images). Each image patch is reshaped into a 64- or 256-dimensional vector for grayscale patches, and a 192- or 768-dimensional vector for color patches. Prior to patch extraction, all images are normalized to have zero mean and unit variance. Any blank patches resulting from random sampling are discarded. The extracted image patch samples are also normalized to zero mean and unit variance to ensure consistency.

\textbf{Resting-State fMRI:} We used the 100 healthy unrelated subject dataset from the WU-Minn Human Connectome Project~\cite{VanEssen13NI}. Each subject was involved in four 15-minute runs with a TR of 0.72 seconds, totaling 1200 frames per run. Standard preprocessing steps were conducted using the Statistical Parametric Mapping (SPM8) package (~\url{https://www.fil.ion.ucl.ac.uk/spm/software/spm8/}) from the Wellcome Institute of Cognitive Neurology, London, including slice timing, realign, coregister, normalize, denoise, and smooth, and the fMRI signal was filtered with a bandpass of [0.01, 0.15]Hz. 

The voxel-level time series were extracted separately for each subject. To reduce the dimensionality of the dataset while preserving the most significant variance, principal component analysis (PCA) was performed at the subject level. This procedure yielded the principal component scores and the percentage of variance explained by each component. We then calculated the cumulative explained variance to determine the minimum number of principal components required to explain at least 99\% of the total variance. To ensure consistency across the selected components, the principal component scores were standardized. This process transformed the scores to have zero mean and unit variance, ensuring equal contribution from each principal component to the analysis. Subsequently, the subject-level data were concatenated, and PCA was applied at the group level to further reduce dimensionality. The PCA-transformed data were used as input for both group-level independent component analysis (GICA) using Infomax and for DDICA to identify independent components. For Infomax, ICASSO~\cite{himberg2003icasso} was run 100 times with a model order of 20 to estimate the most stable and consistent components. The final spatial maps were derived as z-scores, reflecting the absolute strengths of the ICA components from both the group-level ICA and DDICA analyses.

\subsection{Results}
\subsubsection{Benchmarking with Simulated Data}
\hspace{1.2em} In Fig.\ref{fig:simu}, the performance of DDICA is compared to 11 other linear ICA algorithms (as shown in Table.\ref{tab}) in decomposing mixed sources, with the ground truth also shown for reference. Several insights can be drawn. First, DDICA achieves the best performance across all three components when compared to the other linear ICA methods. This is especially clear in components 1 and 2, where DDICA shows excellent separation results that closely match the ground truth. For component 3, although DDICA successfully separates the signal from the mixture, its performance is slightly lower compared to components 1 and 2. Second, among the other 11 linear ICA algorithms, Infomax and FastICA still perform relatively well and outperform many of the remaining methods. In contrast, several algorithms such as Simbec, Amuse, and Evd exhibit visibly weaker separation performance, often failing to recover the shape or structural details of the ground truth components. These visual differences underscore DDICA's ability to maintain spatial fidelity while achieving accurate separation, especially in the presence of noisy mixtures.

\begin{figure}[!ht]
    \centering
    \includegraphics[width=0.48\textwidth,height=8cm]{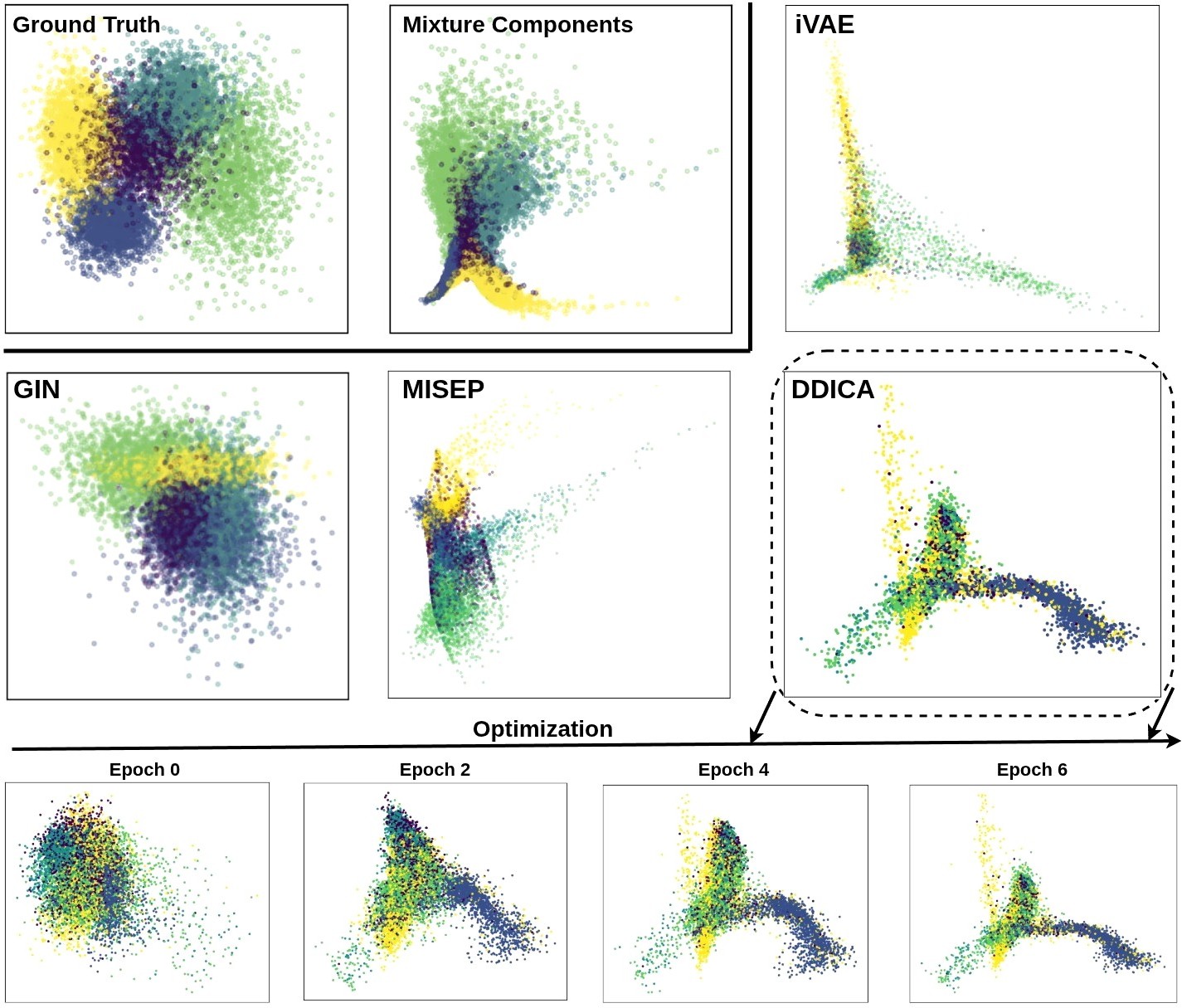}
    \caption{\textbf{Comparison of DDICA with other nonlinear ICA approaches.} The ground truth sources and their corresponding mixture components were generated and visualized. Subsequently, four nonlinear ICA methods-iVAE, GIN, MISEP, and DDICA-were applied to the mixture components to perform source separation. The results of each method are presented separately for comparison. For DDICA, we additionally reported its performance with different stages of optimization.}
    \label{fig:nica10}
\end{figure}

To further quantify the performance of component unmixing, we conducted 90 repeated simulation trials for each linear ICA algorithm. The results are shown in Fig.\ref{fig:simun}, which align with the qualitative observations in the spatial maps. DDICA consistently achieves higher performance compared to the other ICA algorithms across all trials. This suggests that models incorporating joint source information can improve unmixing performance in certain scenarios. 

In addition, Table.\ref{tab} summarizes the mean and standard deviation of spatial correlation scores across trials for each method, further highlighting DDICA’s robustness and stability in recovering independent components.

To further assess the performance of DDICA on highly nonlinear mixed sources, we constructed a particularly challenging dataset by combining three distinct nonlinear signal patterns: double blobs, spiral waves, and circular structures. These components were fused into a single set of complex mixed signals, designed to test the robustness of separation algorithms under severe nonlinear distortions. We then benchmarked DDICA against two of the most widely used alternatives, Infomax and FastICA, in these demanding conditions that go well beyond the assumptions of linear mixing.

As shown in Fig.\ref{fig:n2}, DDICA was able to partially and successfully decompose the mixed signals, recovering clear representations of the original double blob, spiral, and circular components. In contrast, both Infomax and FastICA struggled to isolate the nonlinear sources, producing results with poor structural fidelity and minimal resemblance to the original components. These findings highlight DDICA's enhanced capability in dealing with intricate nonlinear mixtures, where conventional ICA techniques tend to fail.

To further evaluate the performance of DDICA, we compared it with three other deep neural network-based nonlinear ICA approaches: GIN~\cite{Sorrenson2020Disentanglement}, iVAE~\cite{khemakhem20a}, and MISEP~\cite{Almeida02}. The performance of GIN, iVAE, MISEP, and DDICA in separating mixture components is shown in Fig.\ref{fig:nica10}. The results indicate that DDICA achieves performance comparable to the other methods. Additionally, we present snapshots of the separated components during optimization. Overall, the results demonstrate that DDICA performs on par with existing nonlinear ICA approaches.

\begin{figure}[!hb]
    \centering
    \includegraphics[width=0.48\textwidth,height=8cm]{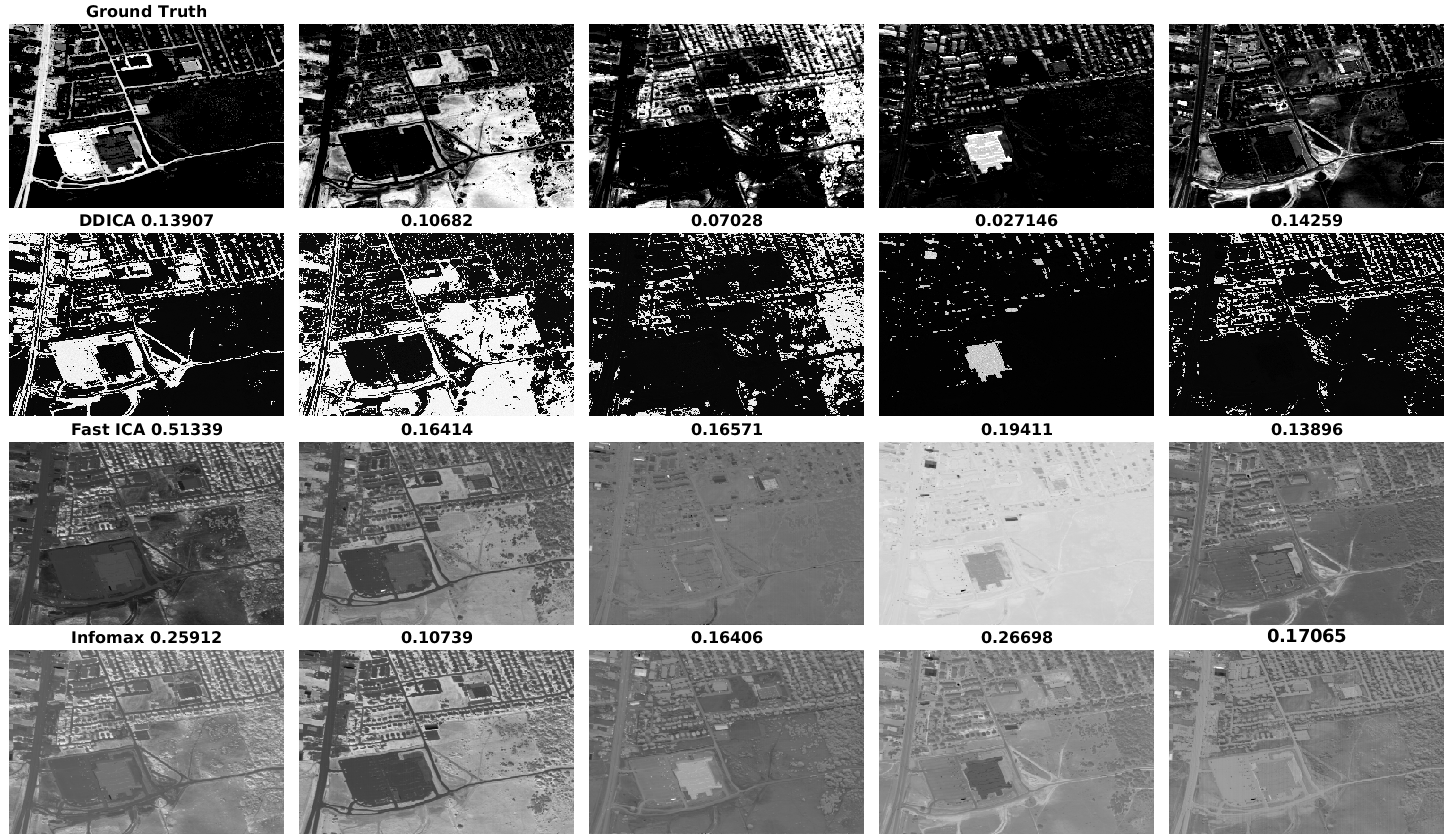}
    \caption{\textbf{Hyperspectral image unmixing.} The first row shows the ground truth components, followed by decomposition results from DDICA (second row), FastICA, and Infomax. To quantify performance, the peak mean squared error (PMSE) between the decomposed components and the ground truth is reported in each component’s title. Lower PMSE values indicate better performance.}
    \label{fig:hyper}
\end{figure}

\begin{figure*}[!ht]
    \centering
    \includegraphics[width=\textwidth, height=13.5cm]{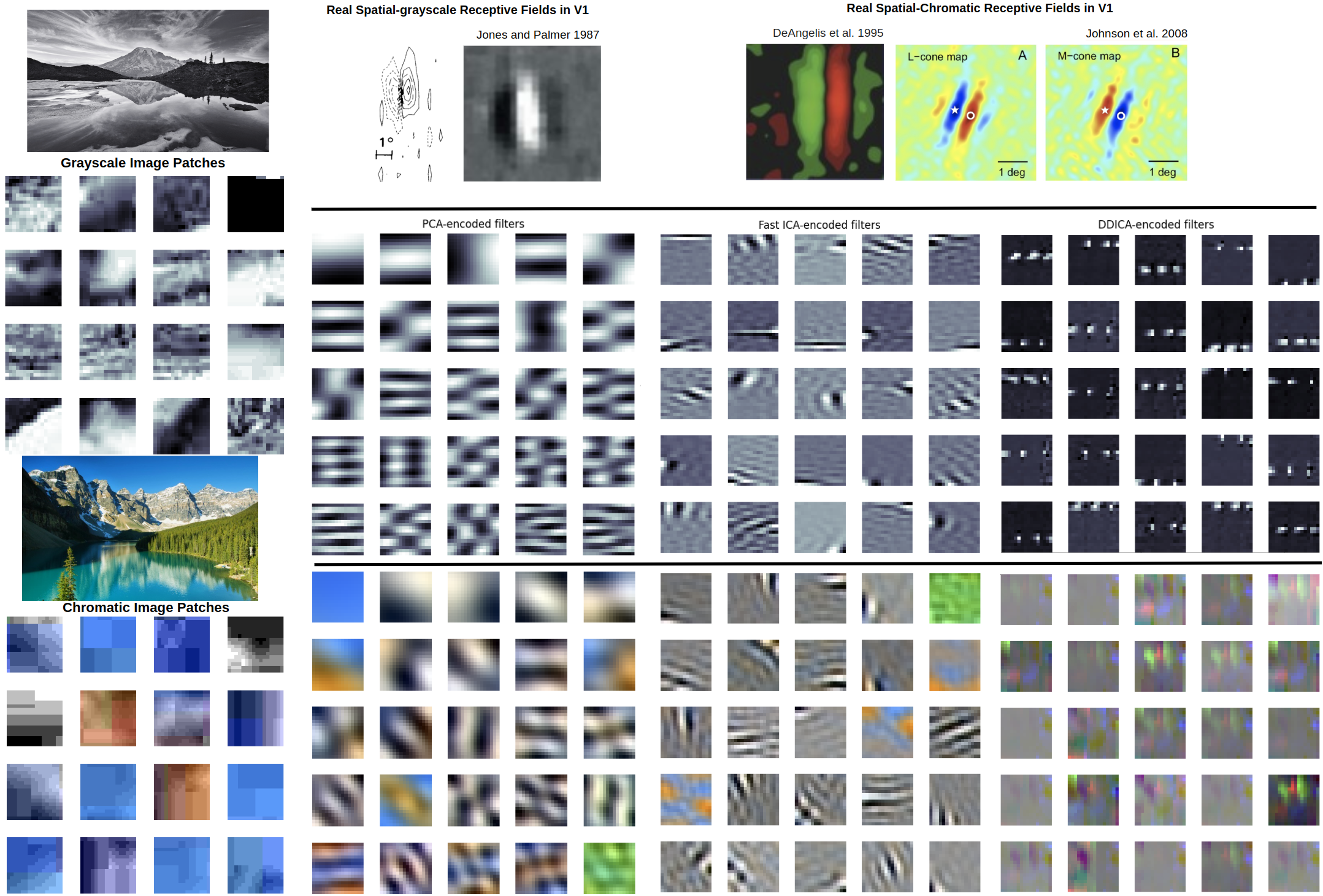}
    \caption{\textbf{Modeling primary visual receptive fields.} DDICA was applied to model achromatic and chromatic primary visual receptive fields and compared with PCA and FastICA. Image patches were first extracted from both achromatic and chromatic stimuli. Encoder filters were then learned using PCA, FastICA, and DDICA, respectively. To evaluate the quality of the learned filters, biologically measured spatial grayscale~\cite{jones1987gabor} and spatial chromatic~\cite{DEANGELIS1995451,johnson2008orientation} receptive fields in V1 are also shown for reference.}
    \label{fig:rpvisual}
\end{figure*}

\subsubsection{Application to Hyperspectral Images}
\hspace{1.2em} In Fig.\ref{fig:hyper}, we present the hyperspectral image unmixing results obtained by DDICA, FastICA, and Infomax, alongside the ground truth for comparison. As observed, DDICA achieved the best overall performance in separating the hyperspectral components, outperforming both FastICA and Infomax. To further quantify the unmixing accuracy, we computed the Peak Mean Squared Error (PMSE) for each decomposed component relative to the ground truth (i.e., DDICA vs. FastICA/Infomax: 0.139 vs. 0.513/0.259; 0.107 vs. 0.164/0.107;  0.070 vs. 0.166/0.164; 0.027 vs. 0.194/0.267; 0.143 vs. 0.139/0.171). The results show that DDICA consistently yielded lower PMSE values, indicating superior decomposition quality compared to the other two methods.

\subsubsection{Application to Modeling Primary Visual Receptive Fields}
\hspace{1.2em} ICA has been used to model primary visual receptive fields, demonstrating that the learned ICA filters exhibit notable statistical similarities to biologically observed receptive fields~\cite{Hateren99}. This supports the idea that ICA can serve as a model for efficient neural coding objectives~\cite{Hyvarinen2009}. In our study, we used sensory input data consisting of both grayscale and chromatic natural images. We then performed a visual comparison between the ICA-derived filters and experimentally measured receptive fields from the visual cortex, as illustrated in Fig.\ref{fig:rpvisual}. The reference receptive fields are drawn from prior neuroscience studies that measured neural activity in the primary visual cortex. For grayscale images, ICA recovered filters that closely resemble the receptive fields of simple cells in the visual cortex~\cite{jones1987gabor}. For color images, ICA produced similar 2D Gabor-like filters, additionally capturing red-green opponency patterns observed in earlier physiological studies~\cite{DEANGELIS1995451,johnson2008orientation}.

\begin{figure}[!ht]
    \centering
    \includegraphics[width=0.5\textwidth,height=4.3cm]{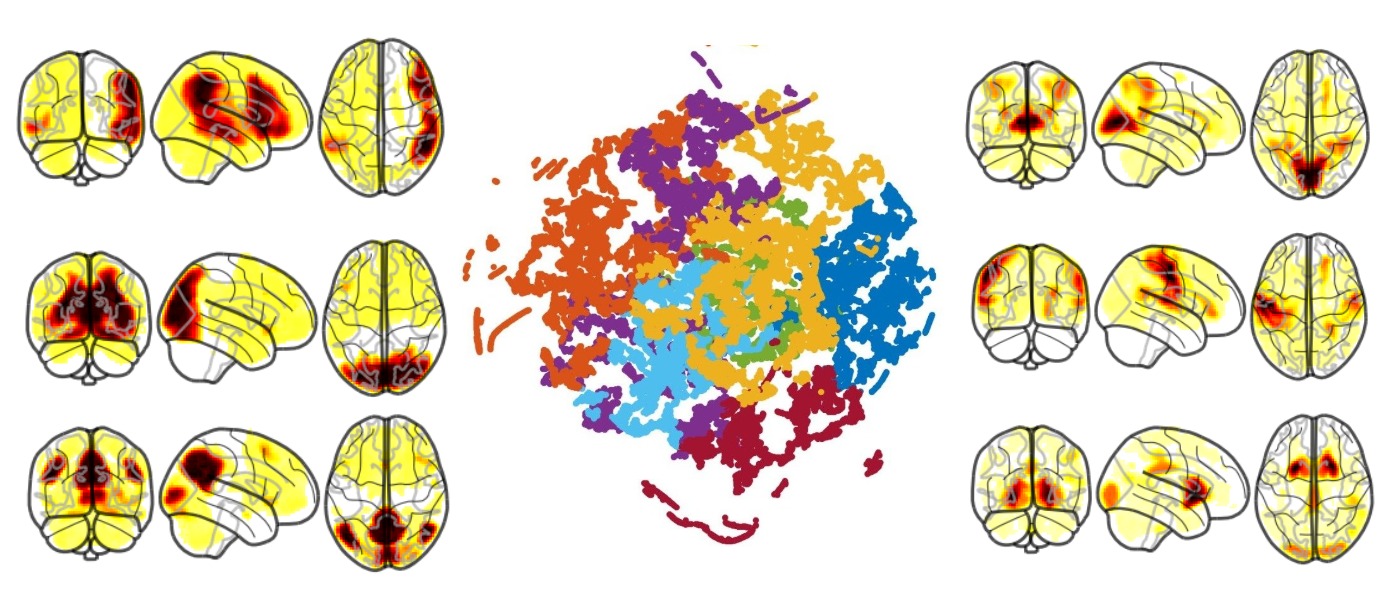}
    \caption{\textbf{Maximally independent brain components estimated from resting-state fMRI.} Six independent brain components were extracted using DDICA. These maps were then projected using t-SNE to reveal and analyze the distribution among the learned components. On the side, spatial maps of the DDICA-derived ICA components corresponding to six clusters are also displayed for visual comparison.}
    \label{fig:ddica_latent}
\end{figure}

We applied DDICA and FastICA to natural grayscale and chromatic images and analyzed the resulting filters. A visual comparison with experimentally measured physiological receptive fields revealed that the filters derived from DDICA qualitatively resemble biological receptive fields, particularly in the case of chromatic responses, where DDICA outperformed FastICA. To further assess DDICA's performance, we also compared its filters to those obtained from a PCA-based encoder. Unlike DDICA, PCA failed to produce filter structures that align with known physiological receptive fields, despite its capacity for input decorrelation. 

These findings suggest that DDICA is more effective than PCA at producing biologically plausible neural codes, and it also outperforms FastICA in modeling chromatic receptive fields in the primary visual cortex.

\subsubsection{Application to Functional MRI}
\hspace{1.2em} To gain deeper insight into the behavior of DDICA, we applied it to resting fMRI data. To better understand how DDICA separates components and encodes information, we further investigated the latent space (maximally spatially independent components) learned by the model. This latent space, which offers a compressed and abstract representation of the fMRI signals, more effectively captures the underlying structure of brain networks identified by DDICA, especially when compared to the performance of the widely used Infomax algorithm~\cite{bell1995information}.

\begin{figure*}[!ht]
    \centering
    \includegraphics[width=\textwidth,height=7.5cm]{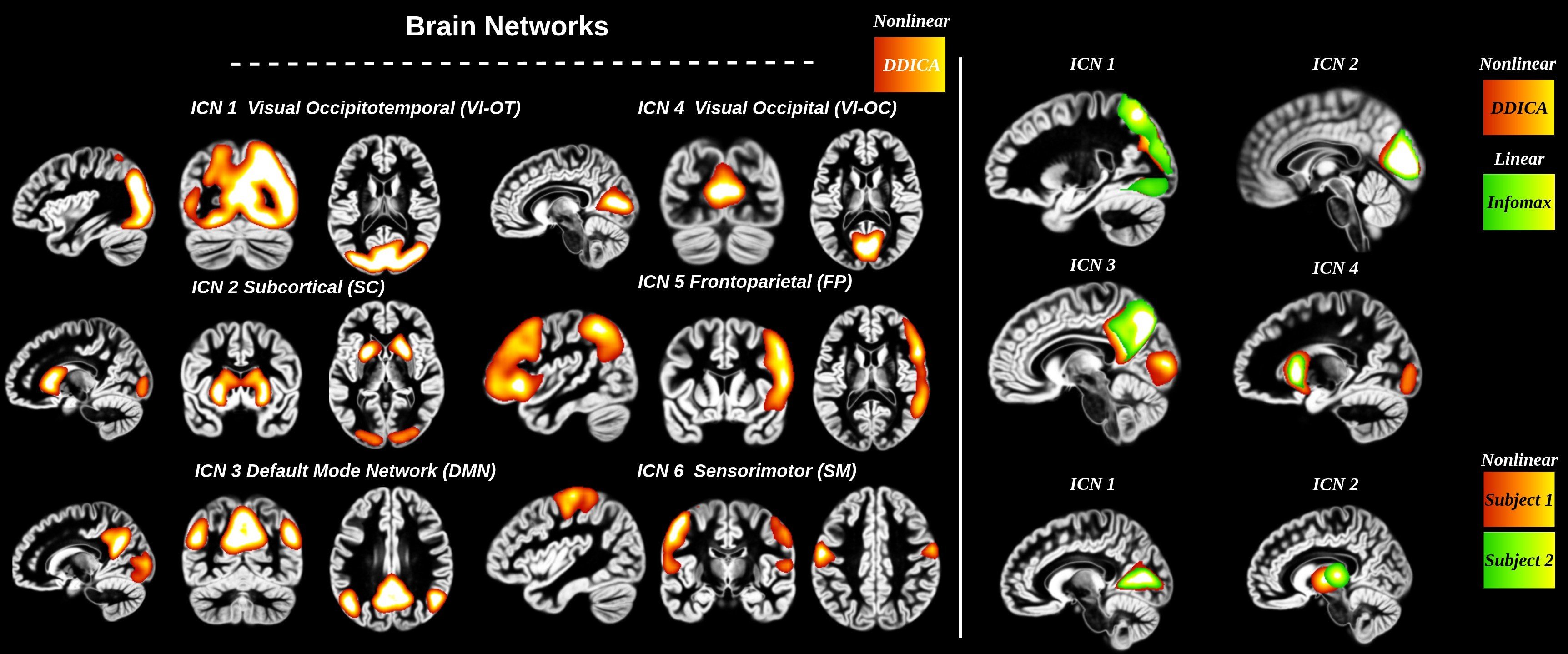}
    \caption{\textbf{Independent networks estimated from resting-state fMRI.} Six canonical brain networks (i.e., VI-OT, VI-OC, SC, FP, DMN, and SM) were presented using DDICA. To compare DDICA with a commonly used linear ICA method (e.g., Infomax) for identifying independent networks from fMRI data, we selected four networks estimated by DDICA (shown in orange gradient) and Infomax (shown in green gradient), respectively. Subject-level spatial network maps were also evaluated to assess the stability of DDICA.}
    \label{fig:fmri}
\end{figure*}

To visualize and further analyze the maximally independent networks from DDICA and group-level ICA (Infomax, ICASSO), we applied t-Distributed Stochastic Neighbor Embedding (t-SNE) to effectively illustrate the distribution of different maximally independent brain networks. In t-SNE, we used the Barnes-Hut approximation for efficient computation~\cite{RUKHSAR}. We utilized Euclidean distance to measure similarity between data points, and the dimensionality was reduced to three dimensions. This approach helps visualize complex high-dimensional data in a more interpretable three-dimensional space, as shown in Fig.\ref{fig:ddica_latent}. By reducing the dimensionality of the learned independent networks, t-SNE illuminated the organization and separation of distinct brain networks identified by DDICA. This visualization offered valuable insights into the functional cohesion of the extracted components while also showcasing the spatial relationships and distinctions among various brain networks.

Secondly, the resting-state networks were extracted using both DDICA and group-level ICA (Infomax, ICASSO). DDICA was used to decompose the fMRI data into latent spatial components, revealing underlying brain networks, as mentioned above. Concurrently, group ICA was performed using GIFT (\url{https://trendscenter.org/software/gift/}) to extract spatially independent components from the resting-state fMRI data. In Fig.\ref{fig:fmri}, six canonical brain networks estimated via DDICA were selected and presented: visual occipitotemporal (VI-OT), visual occipital (VI-OC), subcortical (SC), frontoparietal (FP), default mode network (DMN), and sensorimotor (SM) networks.

To further evaluate the differences between DDICA and Infomax, we selected four components that illustrate both similarities and differences between the two methods. For example, ICN1 and ICN2 represent components that are either the same or have substantial overlap between DDICA and Infomax. In contrast, ICN3 and ICN4 demonstrate that DDICA not only captures components also identified by Infomax, but also reveals additional network features not detected by Infomax, as shown in Fig.\ref{fig:fmri}. Moreover, we also assessed subject-level spatial networks to evaluate the consistency of DDICA across individuals. The results demonstrate that DDICA produces stable and spatially coherent components across subjects, indicating its robustness in capturing individual-level network structures.

\section{Discussion}
\hspace{1.2em} In this study, we present DDICA, a novel method based on deep neural networks that minimizes total correlation using a matrix-based entropy function to enforce statistical independence among components. To validate its effectiveness, we first tested it on two simulated cases. In the first case, we compared its source separation performance against 11 existing ICA algorithms. The results demonstrated DDICA's effectiveness and superior performance. To further assess its ability to handle extremely nonlinear and heavily mixed sources, we generated synthetic data with controllable nonlinearity levels. DDICA was able to partially decompose these complex mixtures, highlighting its robustness in challenging conditions. Additionally, we evaluated DDICA against other nonlinear ICA approaches and found that it achieved performance on par with existing methods. We then applied DDICA to real-world signal separation tasks. One such application involved unmixing hyperspectral images, where DDICA achieved better performance compared to traditional ICA methods. Additionally, we used DDICA to model primary visual receptive fields. Previous studies have shown that filters learned through ICA resemble biological visual receptive fields~\cite{Hateren99,Hyvarinen2009}. Similarly, our results showed that the encoder filters learned by DDICA also exhibit biologically plausible receptive field patterns. Finally, we applied DDICA to resting-state fMRI data, where it successfully identified distinct functional brain networks. Overall, our experiments demonstrate that DDICA is a powerful and generalizable framework for independent component analysis across a variety of data modalities.

While the matrix-based rényi entropy functional offers a powerful nonparametric approach for estimating information-theoretic quantities directly from data, it comes with several limitations. A primary challenge is its poor scalability to large datasets due to the need to compute and store an \( n \times n \) kernel matrix, which imposes significant memory and computational burdens~\cite{Gong22}. Additionally, the method is highly sensitive to the choice of kernel function and its associated hyperparameters, such as the bandwidth in Gaussian kernels, making it difficult to tune effectively across different applications~\cite{Yu2019,giraldo2014measures}. The entropy estimates may also become unstable or biased with small sample sizes, and the approach generally assumes that data samples are independently and identically distributed, which may not hold in structured data~\cite{cover1991information}. Furthermore, the interpretation of entropy values lacks the intuitive probabilistic meaning found in traditional density-based methods, and the dependency on the Rényi entropy order parameter \( \alpha \) can influence results in ways that are not always straightforward. These limitations must be carefully considered when applying the method, especially in large-scale or highly structured data scenarios.

Moreover, a key limitation of DDICA lies in its limited ability to effectively separate complex, heavily mixed sources, particularly in cases involving strong nonlinearity or overlapping components. While the method has shown promising results in separating real-valued signals, it still struggles when applied to scenarios involving complex-valued signals. This limitation becomes particularly significant when considering practical, real-world applications where the observed signals are often not purely real but can include complex-valued components, such as in many communication systems, and biomedical signal processing. The presence of complex signals introduces additional challenges in terms of model assumptions, statistical independence, and the underlying mathematical framework required for accurate source separation~\cite{Aapo23PATTERN,Novey2008ComplexICA}. As a result, more research and methodological advancements are needed to extend the capabilities of DDICA to accommodate these more complex and realistic signal conditions.

Furthermore, DDICA continues to face significant challenges when dealing with extremely nonlinear mixed signals. While DDICA is partially successful in separating these highly nonlinear mixtures, the results are still far from perfect. Specifically, although some level of source separation is achieved, the output lacks completeness, and certain structures and fine details present in the ground truth signals are either distorted or entirely missing. This indicates that the current model struggles to fully capture and reconstruct the complex nonlinear relationships inherent in such signals, highlighting the need for further refinement of the algorithm to improve its performance in highly nonlinear scenarios.

Additionally, when applying DDICA to model primary visual receptive fields using spatial grayscale images as input, we observed that several challenges still remain. Specifically, the results obtained from DDICA do not fully align with the characteristics of real biological grayscale receptive fields. Although DDICA demonstrates strong performance in modeling chromatic receptive fields, particularly in the red-green channel, it fails to accurately capture the blue-yellow channel responses. This discrepancy suggests that, while the current architecture of DDICA is capable of extracting certain biologically relevant features, it still lacks the structural depth and flexibility needed to fully replicate the complexity of the visual system. Therefore, further improvements and refinements in the design and architecture of DDICA are necessary to bridge this gap and enhance its biological plausibility and representational power.

Finally, for resting-state fMRI, future extension work could focus on jointly analyzing both the spatial domain and the corresponding time series. This would allow for a more comprehensive estimation of functional connectivity, which in turn can enhance our understanding of brain cognitive functions and support the exploration of various brain disorders. Furthermore, once well-defined spatial networks have been estimated by DDICA, spatially constrained ICA~\cite{Yuhui20} can be applied to refine the analysis and enhance interpretability.

\section{Conclusion}
\hspace{1.2em} In this paper, we introduced DDICA, an efficient and scalable nonlinear ICA framework designed to extract independent components from complex, and potentially nonlinear, mixed sources. Unlike traditional ICA methods, DDICA leverages a matrix-based Rényi’s $\alpha$-order entropy functional to directly optimize the independence criterion using deep networks, thereby overcoming challenges such as limited flexibility in nonlinear settings. We validated the effectiveness of DDICA through extensive experiments on synthetic datasets, demonstrating its robustness and reliability in recovering latent source signals. In addition, we applied DDICA to three challenging real-world tasks: hyperspectral image unmixing, natural image analysis, and resting-state fMRI. Across all domains, DDICA consistently outperformed conventional ICA algorithms and produced results that are comparable to, or better than, other state-of-the-art ICA-based methods. These findings underscore DDICA's versatility and potential as a general-purpose solution for blind source separation in both linear and nonlinear signal processing applications. 

\section{Data and code availability}
All data and code will be released on \url{https://github.com/qianglisinoeusa/DDICA-Nonlinear-ICA} upon the publication of the paper.

\section{Declaration of Competing Interest}
\hspace{1.2em} The authors declare that they have no known competing financial interests or personal relationships that could have appeared to influence the work reported in this paper.

\section{Acknowledgements}
\hspace{1.2em} This work was supported by the grants NSF 2112455, NSF 2316420, NSF 2316421, NIH R01MH123610, and NIH R01MH119251.
\bibliographystyle{ieeetr}
\bibliography{references} 
\end{document}